\documentclass[iop]{emulateapj}
\usepackage{graphicx}
\usepackage{natbib}
\usepackage{amsmath}

\providecommand{\e}[1]{\ensuremath{\times 10^{#1}}}

\def\lesssim{\mathrel{\hbox{\rlap{\hbox{\lower3pt\hbox{$\sim$}}}\hbox{\raise2pt\hbox{$<$}}}}}
\def\gtrsim{\mathrel{\hbox{\rlap{\hbox{\lower3pt\hbox{$\sim$}}}\hbox{\raise2pt\hbox{$>$}}}}}

\usepackage{hyperref}

\begin{document}
\title{Kepler Flares II: The Temporal Morphology of White-Light Flares on GJ 1243}

\author{
	James R. A. Davenport\altaffilmark{1,2},
	Suzanne L. Hawley\altaffilmark{2}, 
	Leslie Hebb\altaffilmark{3},
	John P. Wisniewski\altaffilmark{4},
	Adam F. Kowalski\altaffilmark{5},\\
	Emily C. Johnson\altaffilmark{2}, 
	Michael Malatesta\altaffilmark{4},
	Jesus Peraza\altaffilmark{2}, 
	Marcus Keil\altaffilmark{4}, 
	Steven M. Silverberg\altaffilmark{4},\\
	Tiffany C. Jansen\altaffilmark{2}, 
	Matthew S. Scheffler\altaffilmark{4},
	Jodi R. Berdis\altaffilmark{4},
	Daniel M. Larsen\altaffilmark{2},
	Eric J. Hilton\altaffilmark{6}					
	}

\shorttitle{Temporal Morphology of White-Light Flares}
\shortauthors{Davenport et al.}
\submitted{} 
\accepted{2014 November 12}

\altaffiltext{1}{Corresponding author: jrad@astro.washington.edu}
\altaffiltext{2}{Department of Astronomy, University of Washington, Box 351580, Seattle, WA 98195, USA}
\altaffiltext{3}{Department of Physics, Hobart and William Smith Colleges, 300 Pulteney Street, Geneva, NY 14456}1
\altaffiltext{4}{HL Dodge Department of Physics \& Astronomy, University of Oklahoma, 440 W Brooks Street, Norman, OK 73019, USA}
\altaffiltext{5}{NASA Goddard Space Flight Center, Code 671, Greenbelt, MD 20771, USA}
\altaffiltext{6}{Universe Sandbox, 911 E. Pike Street \#333, Seattle, WA 98122}

\begin{abstract}
We present the largest sample of flares ever compiled for a single M dwarf, the active M4 star GJ 1243. Over 6100 individual flare events, with energies ranging from $10^{29}$ to $10^{33}$ erg, are found in 11 months of 1-minute cadence data from Kepler. This sample is unique for its completeness and dynamic range. We have developed automated tools for finding flares in short-cadence Kepler light curves, and performed extensive validation and classification of the sample by eye. From this pristine sample of flares we generate a median flare template. This template shows that two exponential cooling phases are present during the white-light flare decay, providing fundamental constraints for  models of flare physics. The template is also used as a basis function to decompose complex multi-peaked flares, allowing us to study the energy distribution of these events. Only a small number of flare events are not well fit by our template. We find that complex, multi-peaked flares occur in over 80\% of flares with a duration of 50 minutes or greater. The underlying distribution of flare durations for events 10 minutes and longer appears to follow a broken power law. Our results support the idea that sympathetic flaring may be responsible for some complex flare events.
\end{abstract}

\section{Introduction}
M dwarfs have long been known for their magnetic activity, most famously in the form of powerful and frequent UV Ceti-type flares. These explosive events are thought to be analogs to the flares we observe on the Sun, but with much larger energies and higher occurrence rates. On the Sun, flares form as the result of violent magnetic reconnection events, and are ultimately a byproduct of the solar magnetic dynamo that forms from the shearing interface between the radiative core and convective envelope. The greater power and frequency with which M dwarfs flare may be due to their turbulent magnetic dynamos. Whether the creation or cooling mechanisms of M dwarf flares are truly the same as on the Sun is unknown. 

A primary limitation of observational flare studies has traditionally been the challenge in gathering statistically complete samples of flares with detailed light curves. Since flares occur stochastically, acquiring a large and detailed sample of M dwarf flares from the ground necessitates studying many active stars of similar spectral types over multiple nights \citep[e.g.][]{moffett1974,hiltonthesis}. These laborious monitoring campaigns may produce light curves for a few hundred flares, with varying degrees of completeness between targets.

Automated surveys that repeatedly image large portions of the sky provide an attractive alternative, as they can efficiently yield millions of individual photometric (or even spectroscopic) measurements of M dwarfs in which to search for flares \citep[][]{kowalski2009,hilton2010,davenport2012,berger2013}. 
However, such aggregate studies usually do not provide temporal information for individual flares, nor complete flare rates for individual stars. Instead, these studies rely on the assumption that the characteristics of flares are common amongst large groups of stars.

With the introduction of dedicated space-based monitoring we can overcome many of these observational challenges. The Kepler satellite \citep{borucki2010} provides a nearly ideal platform to build statistically complete samples of stellar flares. With round-the-clock white light monitoring for over 150,000 stars spanning nearly 4 years, and remarkable photometric precision, Kepler is the portent to a new era of statistical completeness in stellar activity studies across the main sequence \citep{basri2010,walkowicz2011}.

The occurrence rates and energy distributions for flares, and so-called ``superflares'', for many late type stars in Kepler has already been studied to some extent \citep{notsu2013}. However, the detailed information content from the events themselves has yet to be realized. Understanding the temporal evolution (``light curve morphology'') of white light flares is a great utility for planet hunting, where precision flare templates helps improve planet detection efficiency. Stellar flares are also interesting astrophysical phenomena in their own respect. The physics of stellar flare heating and radiation has been studied for many decades \citep[e.g.][]{gershberg1973,houdebine1991,hawley1995}. The source of the white light continuum emission in flares, and its relationship to the high energy and emission line behavior during these events, is an active area of research 
\citep{butler1988,slhadleo,hawley2003,kowalski2013}.

In this paper, we expand on the sample of M dwarf flares described in \citet[][hereafter Paper 1]{hawley2014} and use Kepler 1-minute data to develop a statistical understanding of the morphological characteristics of flare light curves. The detailed generation and cooling (decay) of flares is likely to be strongly dependent on the properties of the host star, such as temperature, age, mass, or surface gravity \citep{pettersen1984}. To control for these physical differences we concentrate on a single target, the active M4 star GJ 1243 (Kepler ID \# 09726699). The dramatic level of stellar activity from this star in the Kepler data, manifested in both long-lived starspots and frequent flares, has previously been noted \citep{savanov2011,ramsay2013}. Our boutique analysis of this single M dwarf allows us to compile a very large sample of flares, unprecedented in its completeness, and provides the foundation for studying flares from many targets in the Kepler database.

The outline of our paper is as follows. We build our sample with automatic selection and manual validation in \S2. Using this clean sample of flares, we create a high fidelity flare template in \S3. The flare template provides us a robust way to decompose complex multi-peaked flare events in \S4. In \S5 we perform tests to determine the completeness and limitations of our sample. We discuss the small number of unusual flares that are not fit by our template in \S6. Finally, in \S7 we conclude with a brief discussion of the context and implications of this work.

\section{Flare Sample}
As one of the few highly active mid-M dwarfs known in the Kepler field, GJ 1243 was the focus of several Kepler ``Guest Observer'' campaigns (GO programs 20016, 20028, 20031, 30002, 30021). Due to both the uniqueness of the Kepler data, and the high level of magnetic activity present, this is a benchmark object for future studies of starspots, rotation, and flares. A rapid rotation period of $P=0.592$ days was established from light curve modulations by two starspots in the first release of Kepler data \citep{savanov2011}, and independently discovered from ground-based data by \citet{irwin2011}. We are conducting a parallel study of the long-term evolution of these starspots, as well as recovering the signature of weak differential rotation in this rapidly rotating star (Davenport et al. 2014, in prep). Preliminary flare rate analysis for GJ 1243 was conducted by \citet{ramsay2013}, and a detailed investigation comparing the flare rates between several Kepler active and inactive M dwarfs is presented in Paper 1.

For this study, we have utilized all 11 months of short cadence data for GJ 1243 available from the primary Kepler mission. We used the latest reduction of the Kepler light curves, which includes the PDC-MAP Bayesian de-trending analysis from \citet{smith2012}. The short cadence data are provided online\footnote{http://archive.stsci.edu/kepler/} in individual months, as opposed to long cadence data, which are provided in whole quarter increments.

\begin{figure}
\centering
\includegraphics[width=3in]{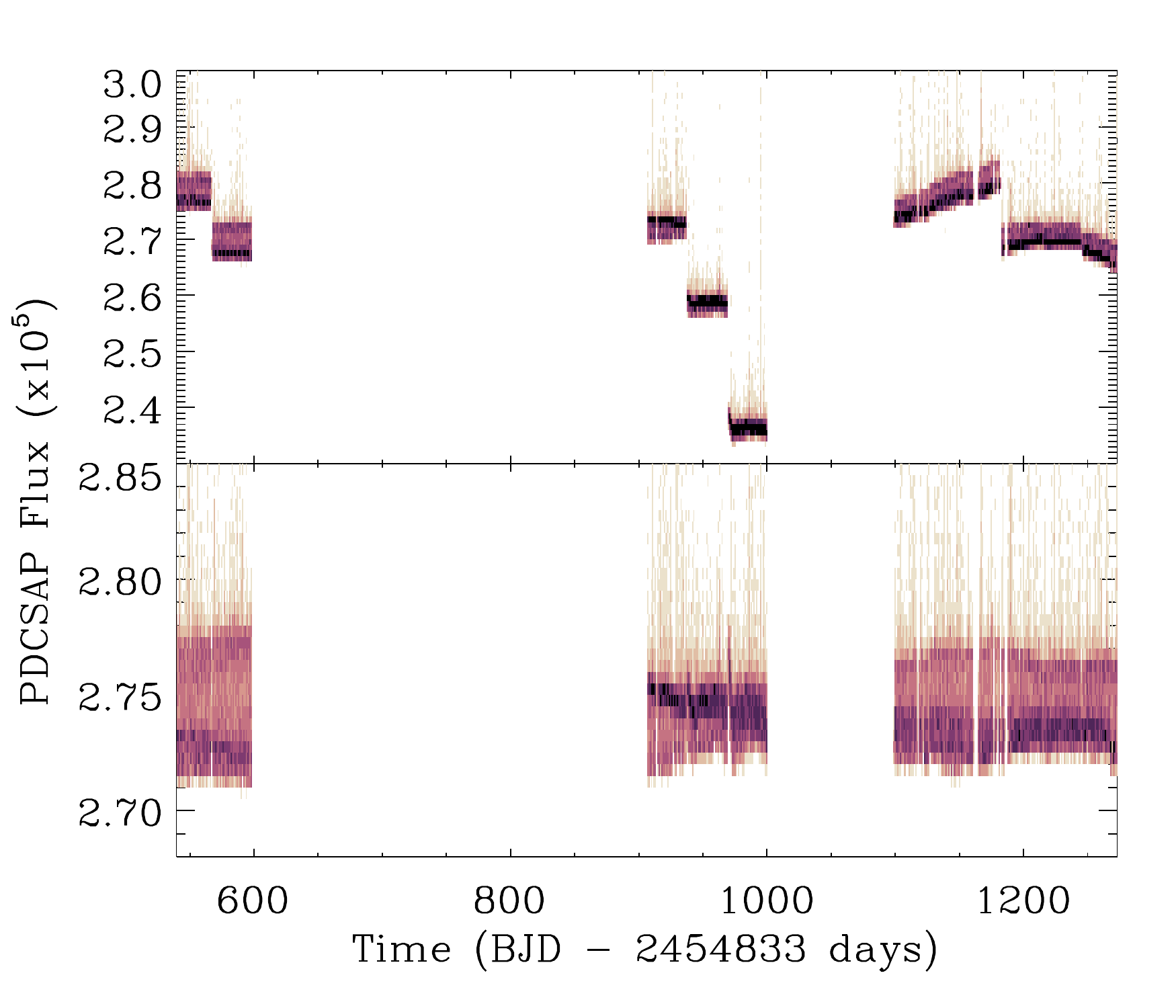}
\caption{Top: Raw PDC-MAP light curve for 11 months of short cadence data for GJ 1243. These data span Quarter 6 to Quarter 13. Bottom: Resulting light curve after our additional linear corrections. Density of points is represented by pixel shade, increasing from light to dark. Note the positive flux excursions due to flares. The median error on the photometry is 78 counts sec$^{-1}$, or $\sigma_F/\bar{F} = 2.9\e{-4}$.}
\label{fig:lc}
\end{figure}

The raw PDC-MAP light curve is shown in the top panel of Figure \ref{fig:lc}. The frequent flares are apparent, visible as positive flux excursions throughout the data. Month-to-month discontinuities in the light curve are present. The smooth trends in the light curve over many quarters, after Time$\approx1100$ days, are suggestive of potentially real changes in the total luminosity of the star on $>100$ day timescales. However, the trends may also be due to errors in the de-trending and calibration, and we speculate that the real starspot and flare variability may be the primary cause for these errors. We subtracted a linear fit from the light curve of each individual data-month to remove any long term calibration errors. The resulting ``flat'' light curve is shown in the bottom panel of Figure \ref{fig:lc}.

We next describe the process to select a reliable sample of flares from the light curve. An iterative approach was used, first employing an automated flare-finding method, followed by a manual (``by-eye'') validation and classification for every flare. The final selection of flare events represents agreement between many such by-eye validations. 

\subsection{Automatic Selection}
The first step in our flare selection from the short cadence light curve was an unsupervised detection of candidate flare events. The starspots produce $\sim$3\% flux variations, which are smooth sinusoidal features that evolve slowly (timescales greater than $\sim$100 days). A great many flares exist in the light curve with amplitudes smaller than the starspot, and as such a simple flux threshold selection would only be useful in detecting the largest flares.

Instead, we subtracted the starspot features using a custom smoothing function. The smoothing function used a variable span smoothing method, inspired by the Supersmoother algorithm \citep{supersmoother}, and with a three-pass iterative approach. The light curve was first smoothed with a large boxcar kernel of 75 data points. All data falling more than 1$\sigma$ away from the boxcar smoothed light curve were removed, and the entire light curve was then fit using a cubic spline. This process was repeated twice more, using progressively smaller boxcar kernels. The resulting smoothed model light curve was only minimally affected by the presence of large flares, and effectively traced the starspot. The starspot curve was then subtracted from the original light curve. We emphasize that while this iterative ``sigma-clipping'' method is not always the favored approach, it provided a rapid and effective means to {\it identify} flares, whose properties were then measured on the original data.

\begin{figure*}[!th]
\centering
\includegraphics[width=5.6in]{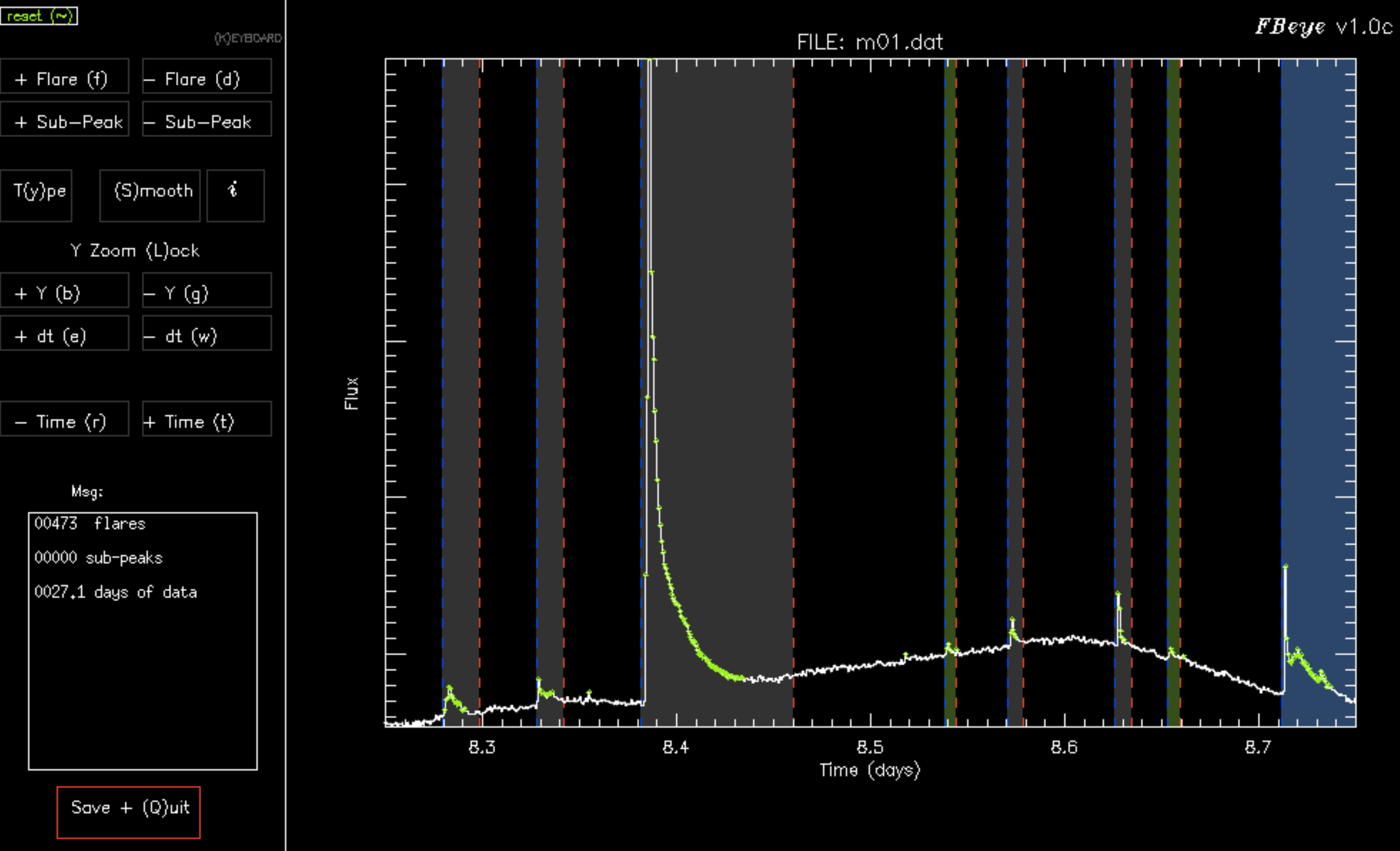}
\caption{Screenshot of our interactive flare-finding suite, FBEYE. User controls to manipulate the light curve display and tag and classify flares are presented in the panel on the left. One half day window of the light curve for GJ 1243 is shown on the right. Displayed are the flares for this time window identified by the auto-finding method detailed in \S2.1. The flare start times (blue dashed lines) and stop times (red dashed lines) define the candidate flare events. Epochs that the starspot removal algorithm identified as positive 2.5$\sigma$ flux outliers are marked with green diamonds. The background shading behind each flare indicates one of the four possible flare type classifications: ``classical'' (grey), ``complex'' (blue), ``unusual'' (pink, not shown), and ``maybe'' (green).}
\label{fig:screenshot}
\end{figure*}

From the flattened (starspot-free) data, we selected all single epochs with positive flux excursions greater than 2.5 times the standard deviation ($\sigma$) of the light curve. Examples of these epochs are highlighted in Figure \ref{fig:screenshot} as green diamonds. Flare candidates were required to have at least 2 consecutive epochs that passed this threshold cut. To avoid inadvertently splitting flares with complex morphologies (e.g. multi-peaked events) in to separate events, neighboring candidate events were merged in to a single flare event if they were only separated by one epoch below the 2.5$\sigma$ threshold. We initially found that the recovered flare candidates systematically had a shortened duration compared to those defined by visual inspection of the first month of data examined. This was due to over-smoothing during the starspot removal. A final empirical correction to the end-time of the candidate flares accounted for this truncation, which extended the candidate flare duration by $\sim$57\%. The effect of this correction is evident in the flares shown in Figure \ref{fig:screenshot}, with the event duration extending beyond the epochs highlighted with green diamonds. The completeness of recovering flares with this automatic algorithm is presented in detail in \S5.

A crude flare type classification was also automatically assigned. By default, all flare candidates were considered ``classical'' (one peak) unless they passed some conservative criteria. Candidate events with durations shorter than four minutes were labeled as ``maybe'' flares, indicating uncertainty in their identification. Events with a duration of at least 20 minutes, and having a secondary peak with amplitude greater than 30\% the maximum peak (either before or after the maximum peak) were labeled as ``complex''. This simple type classification was not robust enough for scientific analysis, frequently missing true complex events, but was helpful in the human validation stage.

\subsection{By-Eye Validation with FBEYE}
The sample of automatically detected flare candidates was then validated by manual inspection. To facilitate this validation, we developed a tool called Flares By EYE (hereafter FBEYE), an IDL suite of programs to view light curves and identify and classify flares. FBEYE also contains the auto-finding routines described above. Users interacted with FBEYE using an interactive graphical window, shown in Figure \ref{fig:screenshot}.

Briefly, the workflow for processing data with FBEYE is as follows: Users first load the light curve of interest, typically a single month of short cadence data. FBEYE runs the smoothing and auto-finding algorithms described in \S2.1, and identifies candidate flares. Then, starting at the beginning of the light curve, the user will scan through the entire month of data, inspecting each flare. The user can choose to look at the smoothed, starspot-subtracted light curve, or the original data (shown in Figure \ref{fig:screenshot}. For each time window, the user will validate all the flares present. Flares that have been incorrectly selected are deleted. Events with incorrect start or end times are modified as needed. Flares that were missed entirely by the auto-finder are then defined. Flares with the wrong type classification, e.g. classical versus complex, have the correct classification assigned. The user then steps forward by half-window increments and repeats the validation processes. Every operation with FBEYE is automatically saved, updating the master flare list for the month.

Typical users would validate a single month of GJ 1243 data in 1--2 hours, while the auto-finder takes only a few seconds to run on a typical workstation. This highlights the great expense in human labor needed to validate these events in a dataset as rich as Kepler, and for a star as active as GJ 1243. Each of the 11 months of short cadence GJ 1243 data was inspected by at least five separate users, with a maximum of eight users, producing over 37,000 individual (though not necessarily unique) flare identifications. 

\begin{figure*}[t]
\centering
\includegraphics[width=6.6in]{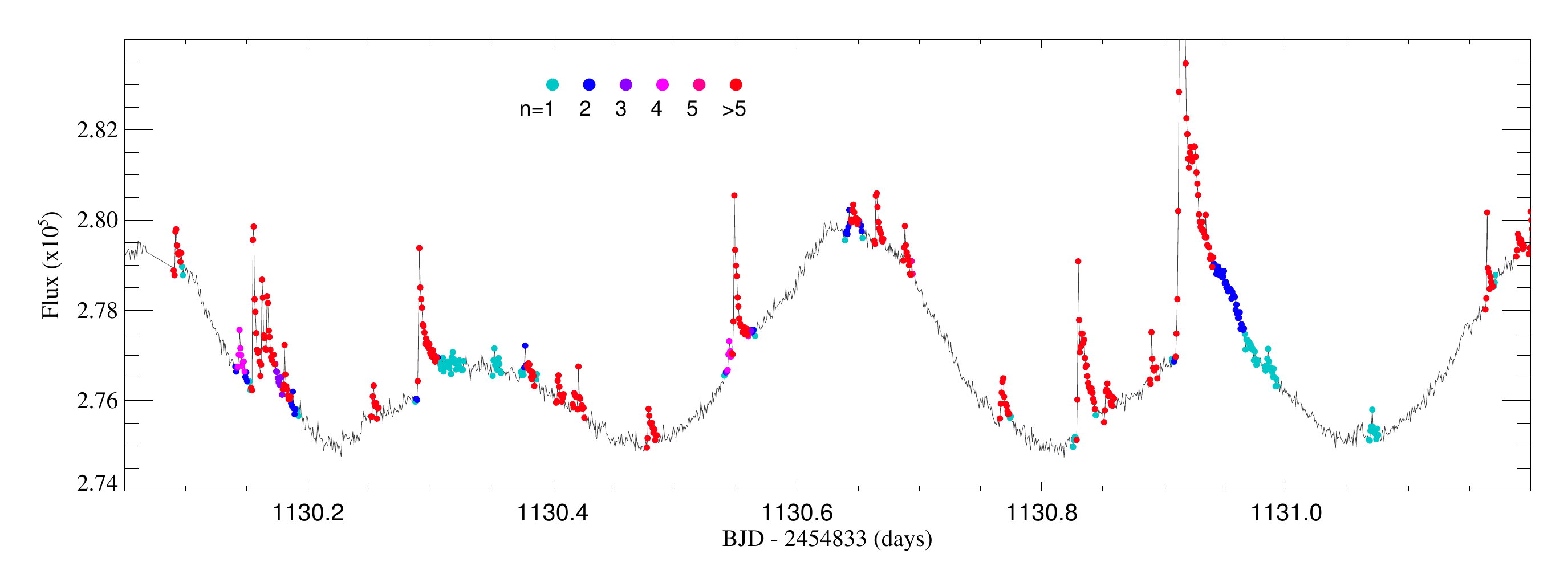}
\caption{A portion of the GJ 1243 light curve from Quarter 12 (February 2012), with epochs that users of FBEYE identified as having flares highlighted. The number of users who identified each epoch is denoted by colors, from blue to red, as indicated in the legend. The end of the gradual decay phase for large flares is the region with the most user disagreement.}
\label{fig:users}
\end{figure*}

Once the data were manually validated, we selected the final flares from a composite of all the user identifications. Figure \ref{fig:users} shows a representative day of data from GJ 1243, with epochs manually selected as containing a flare highlighted, and agreement between users indicated with colors. Many flares show a color gradient, particularly in the decay phase, indicating the regions of strongest disagreement between users. We selected our final flare start and end times to include epochs with at least two users indicating the presence of a flare. As with the auto-finding procedure, flare events were defined as continuous sequences of epochs. Events that were separated by a single non-flaring epoch were merged. The final flare classifications were simplified in to ``classical'' and ``complex''. Events that had been tagged as ``maybe'', either by hand or automatically, where called ``classical'', while the few ``unusual'' events were called ``complex''. The mode of the users' flare classifications was used for each event. The maximum number of users who agreed on each event, as well as the number of users who inspected the corresponding month's data, were also saved in the flare database.

\subsection{Sample Properties}
The final sample of flares from our FBEYE analysis of GJ 1243 contained 6107 unique events. Of these, 5162 were classified as ``classical'', and 945 as ``complex'' (15.5\%). The general properties of these flares, such as the correlation between flare duration and energy from $\sim$1000 flares (2 months of data) are presented in Paper 1. The sample of $>$6000 flares presented here is the largest catalog of stellar flares for a single object (excluding the Sun) that we are aware of, and will be a benchmark for future studies of stellar activity across the main sequence. In Table \ref{datatable}, we present the total number of flares and number of users who examined each month. While the month-to-month number of flares varies considerably, no overall trend in the flare rate was observed.

In Paper 1 we found the quiescent luminosity for GJ 1243 to be log L = 30.66 erg s$^{-1}$ in the Kepler bandpass using ground-based spectrophotometry. By integrating the fractional flux under each flare light curve, we computed the equivalent duration \citep[hereafter ED;][]{gershberg1972,huntwalker2012}, which has units of seconds. Multiplying the ED by the quiescent luminosity for GJ 1243, we found that the largest flares in our sample had energies of $\sim$10$^{33}$ erg, while the smallest were $\sim$10$^{27}$ erg.

\begin{deluxetable}{ccccc}
\tablecolumns{5}
\tablecaption{Monthly breakdown statistics for our sample of flares on GJ 1243. Duration is given in units of days.}
\tablehead{
	\colhead{Month \#}&
	\colhead{Quarter} &
	\colhead{Duration} &
	\colhead{\# Users} &
	\colhead{\# Flares}
	}
\startdata
1  & 6a & 27.1 & 8 & 560 \\
2  & 6b & 30.9 & 7 & 680 \\
3  & 10a & 30.1 & 6 & 618\\
4  & 10b & 31.4 & 7 & 594\\
5  & 10c & 30.2 & 7 & 534\\
6  &  12a & 26.5 & 6 & 499 \\
7  & 12b & 27.4 & 7 & 568 \\
8  & 12c & 27.1 & 6 &  455 \\
9  & 13a & 32.3 & 6 &  505 \\
10 & 13b & 29.2 & 5 &  595 \\
11 & 13c & 27.2 & 5 &  499
\enddata
\label{datatable}
\end{deluxetable}


\vspace{0.1in}

\section{Empirical Flare Template}
Stellar flares are believed to share a common underlying formation mechanism. We wish to investigate if the observed flare morphology can be described by a small number of free parameters and a sufficiently accurate model. Such a template may differ between individual stars, due to a possible dependence on properties such as stellar effective temperature or magnetic field strength. We limit our analysis and discussion, however, to this single star where such variations are assumed to be insignificant. Our sample of flares on GJ 1243 provides a unique dataset for studying the fundamental morphology of flares. Most previous studies of flare statistics have focused on specific measurable properties of individual flares, such as amplitude and duration (see Paper 1). We took a different approach, combining our large sample of individual events to produce a single, high-fidelity template. In this section we describe the creation of this template, whose temporal morphology is described using only two free parameters.

We limited the data to the best observed flares, with an estimated total duration of at least 20 minutes, and with a by-eye classification of ``classical''. We also omitted flares with durations greater than 75 minutes,  as these had a higher likelihood of being complex events. This yielded 885 flares for use in our empirical template. We experimented in adjusting this minimum duration limit, using values ranging from 5 minutes to 50 minutes. The final resulting shape of the empirical flare template was insensitive to the choice of this limit.

For each flare we first subtracted the local quiescent flux level using a linear fit between small time windows before and after the flare start and stop times, respectively. This was done using the non-smoothed initial version of the light curve, detailed in \S2, and effectively subtracted the local effect of the starspot modulations. The starspot-subtracted light curve used in flare detection was not used for building our empirical flare model, as the flares themselves (especially large amplitude flares) could skew the local smoothing prescription used to remove the starspot, and thus affect the resulting template flare shape. After subtracting the local continuum, we then divided each flare by the maximum flux within the event, normalizing the flare to have a relative flux range from 0 before and after the event, to 1 at the flare peak. This scaled amplitude is the first free parameter in our template.

Since the decay phase of a flare (defined as the time between the flare's peak and its return to quiescence) dominates the observed timescale, as well as the total flare energy emitted in white light, care must be taken in normalizing the flare timescales to a common range \citep{kowalski2013}. The choice of a timescale factor is not as straight forward as for the flux amplitude, however. The physics involved in the  rise and decay phases are undoubtedly different, and only weakly correlated (Paper 1). This might indeed suggest the need for separate light curve timescales (or ``stretches'') for the impulsive rise, impulsive decay, and slow decay phases, rather than a single timescale for the entire flare event. This is reminiscent of the 1- versus 2-stretch discussion in normalizing supernovae light curves \citep[][]{perlmutter1999,hayden2010}. However, the 1-minute cadence of Kepler was too coarse to effectively study the shape of the very rapid rise-phase of flares \citep{moffett1974,houdebine1991}, with most flares having rise times (start to peak duration) of only a few minutes. As such we used a single timescale, and encourage future flare studies with higher cadence to specifically revisit the detailed morphology of the rise phase.

We measured the light curve full time width at half the maximum flux, denoted $t_{1/2}$, which included both the impulsive rise and decay components. This metric was used in \citet{kowalski2013}, and is the second free parameter in our template. Since the rise and impulsive decay phases are so rapid, we linearly interpolated each flare to a 10X higher time resolution of 0.1 minutes to find a more accurate value of $t_{1/2}$. Each flare was set to a relative timescale, centered at the time of peak flux, and then normalized by the characteristic timescale $t_{1/2}$. We note that $t_{1/2}$ is dominated by the impulsive decay phase of the flare, while the total energy emitted throughout the decay phase (in the Kepler bandpass) is split between the impulsive and gradual decay phases. We emphasize that the choice of $t_{1/2}$ as the time normalization factor assumes that the entire flare is governed by a single characteristic timescale, and therefore the impulsive and gradual decay phases are connected and do not vary independently. We discuss this further in \S7.

\begin{figure}[]
\centering
\includegraphics[width=3.5in]{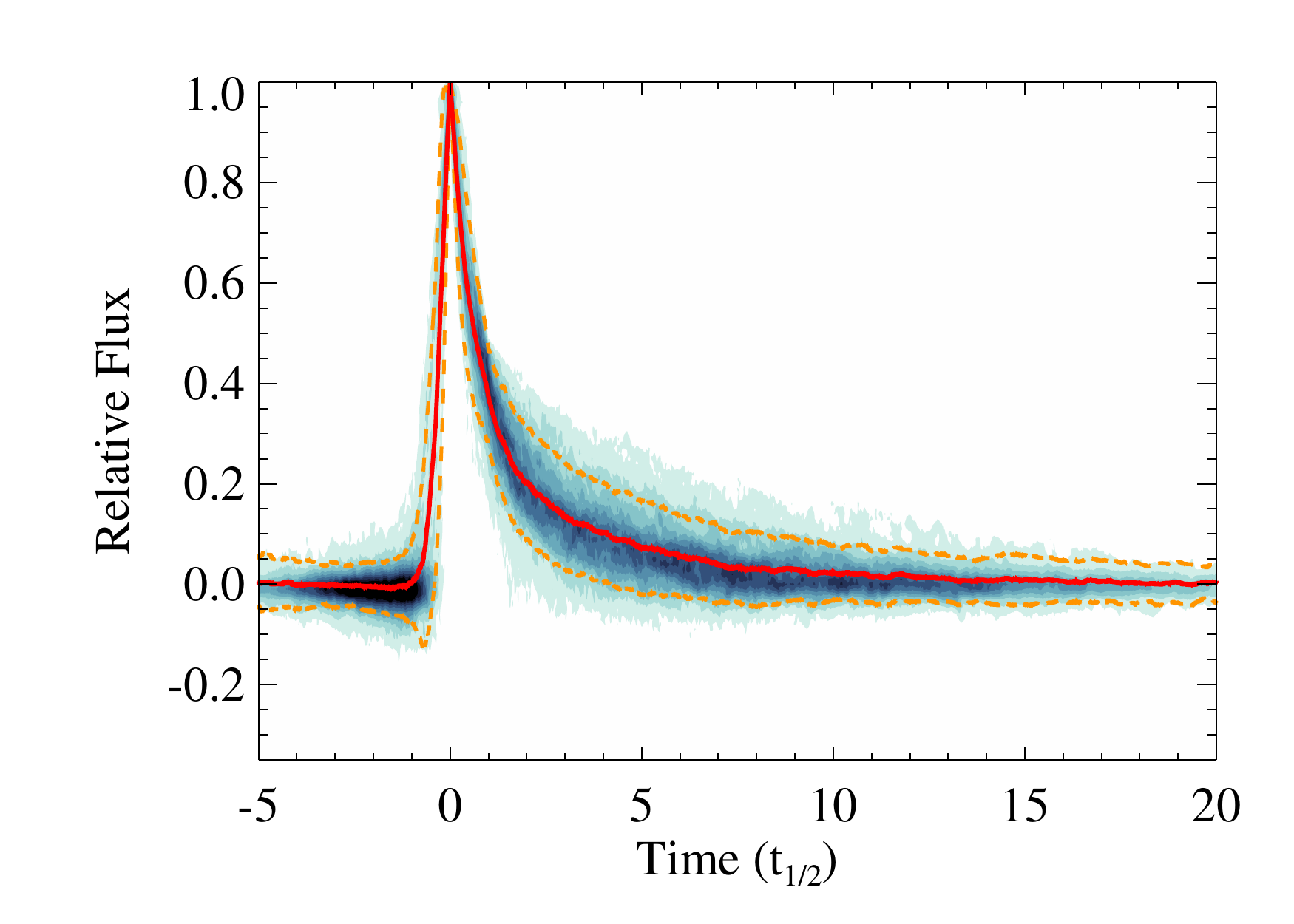}
\caption{Overlay of all 885 classical flares used for the template construction, scaled to relative time and amplitude, and resampled to $\delta t=0.001 t_{1/2}$ time resolution (blue contours). Contour levels increase from light to dark in units of 50. The median of all 885 flares in each time step (red solid line), as well as the robust standard deviation (orange dotted lines), are overlaid. Outlier points are primarily due to errors in local de-trending of the starspot by other nearby flares.}
\label{fig:medflare}
\end{figure}

The entire sample of 885 flares is shown in scaled flux versus scaled time units in Figure \ref{fig:medflare}. Since the sample of flares spans several orders of magnitude in duration, each flare has been resampled to a common relative time resolution. A linear interpolation was used to resample the flares to a time resolution of $\delta t =0.001$, ranging from $t=-5$ to $t=20$ in $t_{1/2}$ scaled time units. We then computed the median flux value for all 885 flares at each $\delta t$ bin, shown as the red line in Figure \ref{fig:medflare}, which defined our fiducial flare template, described by only two free parameters: the amplitude and the scale time ($t_{1/2}$). 
We chose to place the flare peak at $t_{1/2}=0$, so the rise phase occurs in negative time units, and the decay phase in positive time units. We note that a very slight dimming may be visible in the median template before the initial impulsive rise. Pre-flare dimming has been reported by several authors \citep[e.g.][]{hawley1995}, and has been attributed to a temporary elevation of the Balmer continuum absorption in the upper chromosphere, due to an increase in the local electron density caused by ionization from the electron beam \citep{abbett1999,allred2006}. This pre-flare dimming was observed over much shorter timescales than the small effect in our template, and so we can not rule out other phenomena such as filament eruption. However, the possible slight dimming in our flare template could also be due to artifacts in the starspot subtraction.

\begin{figure}[]
\centering
\includegraphics[width=3.5in]{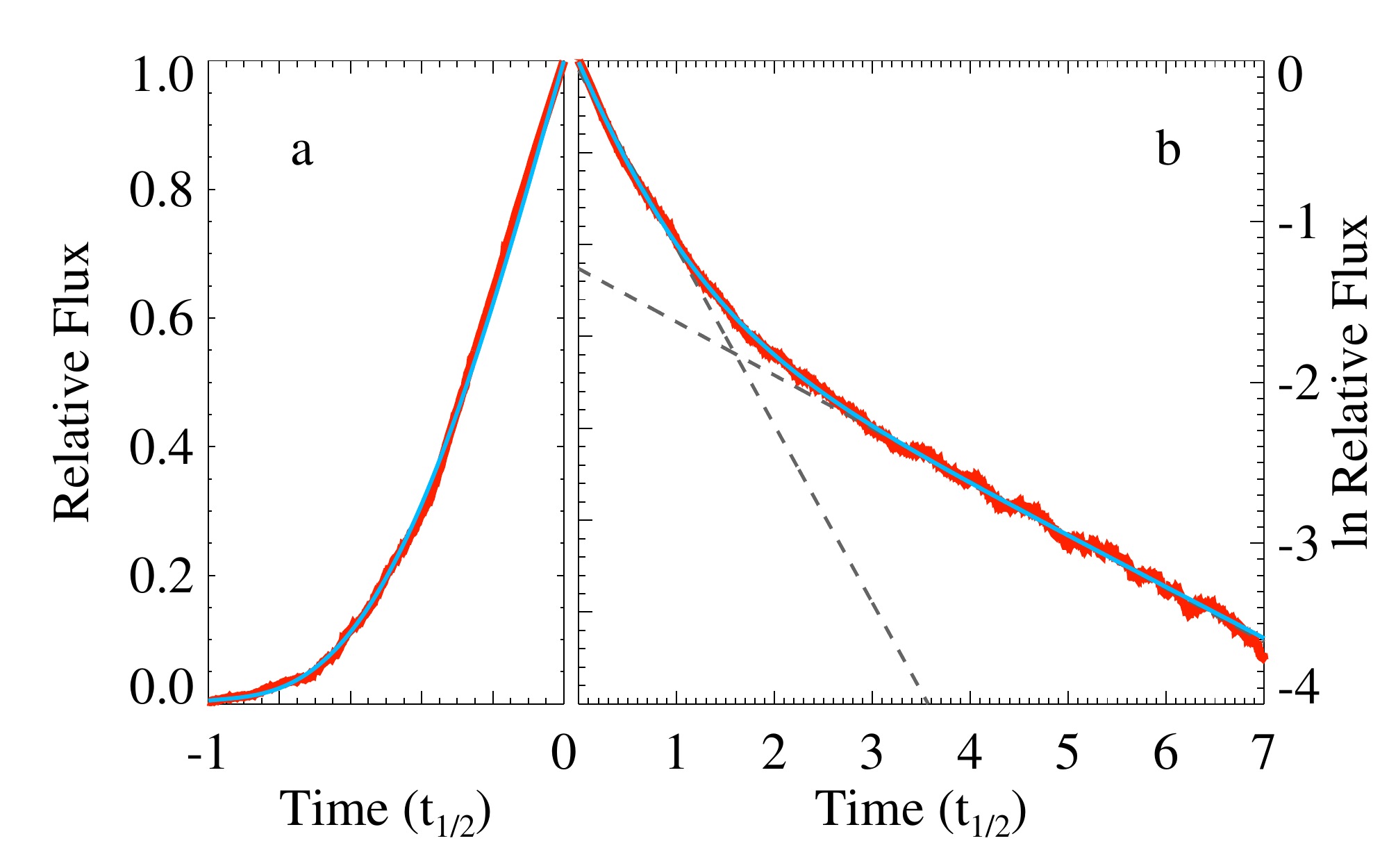}
\caption{{\bf a)} Rise phase of the flare template (red line) in relative flux units, fit with a fourth order polynomial (blue line). The fit was forced to go through relative flux of 1 at time 0. {\bf b)} Decay phase of the flare template (red), in natural logarithm flux units. Straight lines in this space correspond to exponential functions. Two decay regimes are present, and are fit with single exponential functions (grey dashed lines) and a double exponential curve (blue line).}
\label{fig:risedecay}
\end{figure}

\subsection{Analytic Rise Phase Fit}
The median flare template can be described to a very high precision with low-order analytic functions, which in turn makes fitting real flares straight forward. We have divided the flare template into two regimes: the rise and decay phases, or equivalently in Figure \ref{fig:medflare}, $t\le 0$ and $t>0$ respectively. The resulting continuous functions can also be compared with theoretical predictions for flare heating and cooling curves.

The observed rise time for most flares in our sample was between 1 and 5 minutes (see Paper 1), yielding very little information about the morphology of the rise phase with the Kepler 1-minute cadence for any particular flare. However, with our high-fidelity template, details about the rise phase do emerge as shown in Figure \ref{fig:risedecay}a. We fit the rise phase with a 4th order polynomial\footnote{using the IDL non-linear least squares package MPFIT \citep{mpfit}} spanning the time region $-1<t_{1/2}\le0$, chosen to reproduce the initial slow rise and nearly linear behavior near the peak. The fit was forced to go through the point (0,1), in other words to equal a relative flux of 1 at maximum light. The best-fit solution took the form:
\begin{eqnarray}
F_{\rm{rise}} = 1 &+ 1.941(\pm0.008) t_{1/2} -0.175(\pm0.032) t_{1/2}^2  \nonumber \\ 
&-2.246(\pm0.039) t_{1/2}^3 - 1.125(\pm0.016) t_{1/2}^4\, ,
\end{eqnarray}
where values in parentheses indicate uncertainties on each fit coefficient. This analytic model is shown in Figure \ref{fig:risedecay}a, and we emphasize it is valid only for the time region $-1<t_{1/2}\le0$. At times before $t_{1/2}= -1$ the analytic model was forced to equal zero relative flux.

The initially gradual rise, followed by a very rapid climb towards peak flux is similar to the morphology seen in ground-based white light photometry for many of the ``Impulsive Flares'' observed in \citet{kowalski2013}. However, we were not able to resolve any significant ``roll over'' or flattening of the median flare shape as it approached maximum light, as has been observed with much higher time cadence observations \citep[e.g.][]{kowalski2011}. This is due to both the coarse sampling of our light curve compared to the duration of this peak phase, as well as our method of stacking flares based on their observed peak time (the mid-point of the exposure of maximum flux). We believe the latter reason is the primary cause for the ``sharpness'' of our empirical template at maximum light. Our flare model could be used in future studies to provide a robust peak time estimate for flares, and thus investigate the detailed morphology at the light curve maximum with higher time resolution data.

\subsection{Analytic Decay Phase Fit}
Previous efforts to fit the decay phases (often termed ``cooling curves'') for flares have used a variety of parameterizations. A generic exponential decay with time is frequently assumed in searches for flares in large catalogs \citep[][]{walkowicz2011,parkeloyd2014}. A departure from a single exponential (or ``teapot'') cooling curve towards a more detailed two-phase model for large M dwarf flares was explored as early as \citet{andrews1965}, who parameterized the two phases as a steep linear decline, followed by a gradual inverse square shape. The linear regime was then postulated to be due to Bremsstrahlung radiation of fully (or near fully) ionized hydrogen in a small region, while the gradual phase indicated radiative recombination. \citet{hiltonthesis} parameterized the decay phase observationally with an initial linear decline followed by an exponential profile. Recently, \citet{kowalski2013} determined that the impulsive cooling phase in the white light was dominated by the decay of a 10$^4$ K blackbody, which cools rapidly to $\sim$8000 K. A red continuum component, dubbed the ``Conundruum'', is often present during this impulsive cooling phase, and begins to dominate during the gradual cooling phase. Balmer continuum radiation is also observed throughout the flare cooling, but at a lower level.

The decay portion of our median flare template exhibited two clear exponential regimes, seen in Figure \ref{fig:risedecay}b: an initially rapid decay, followed by a longer timescale gradual phase. The flare template is displayed using the natural log of the flux in Figure \ref{fig:risedecay}b, so that straight lines correspond to exponential functions of the form $F(t) = a \, e^{\,(b\, t_{1/2})}$. A single exponential curve is clearly inappropriate in representing the flare decay. The shape was also not well fit using a single power law. 

We fit the decay profile using two separate parameterizations. Our initial model used least-squares minimization to fit the median flare template with two exponential curves. The fit was computed in two well-separated time regions: $0<t_{1/2}< 0.5$ and  $3<t_{1/2}<6$. The resulting exponential fits were:
\begin{eqnarray}
F_1 = & 0.948\, e^{-0.965\, t_{1/2}} \\
F_2 = & 0.322\, e^{-0.290\, t_{1/2}}
\end{eqnarray}
respectively. The two curves intersected at $t_{1/2}=1.60$, and relative flux = 0.20 (ln flux $= -1.595$), where the flare is presumed to instantaneously switch between decay profiles. This corresponds to the time of transition between the impulsive and gradual decay phases, and occurs at nearly the same relative flux value found by \citet{hiltonthesis}. Such an abrupt threshold between the impulsive and gradual phases would represent a state change in the flare cooling, or rapid change between the dominant emission components.

The energy budget of the template flare shape can then be divided into three regimes: the impulsive rise, impulsive decay, and gradual decay, with the transition between decay phases set using the intersection of the two exponential curves defined above. Integrating the flare template within each regime, we found that the rise phase contains 19.9\%, the impulsive decay phase 41.1\%, and the gradual decay phase 38.9\% of the total energy. By combining the impulsive rise and decay phase values, the fraction of total energy emitted during the impulsive phase observed in the Kepler band is 61\%. This is close to the $V$- and $R$-band continuum energy fractions found in a very large flare on AD Leonis \citep[Table 6 of][]{slhadleo}.

However, the template flare decay profile in Figure \ref{fig:risedecay}b indicates a smoother transition between decay phases. We thus fit the entire decay profile of the flare template with a continuous function, using the sum of two exponential curves:
\begin{eqnarray}
F_{\rm{decay}} = &0.6890(\pm 0.0008) \, e^{-1.600 (\pm0.003)\, t_{1/2}} + \nonumber \\
	&0.3030(\pm 0.0009)\, e^{-0.2783 (\pm0.0007)\, t_{1/2}}\,.
\end{eqnarray}
This parameterization would represent two physically distinct regions, each with its own exponential cooling profile, radiating throughout the entire flare decay. The initial decay would then be dominated by a brighter (presumably hotter) region that cools more quickly, and the gradual decay to a cooler region that cools more slowly, similar to the physical description of the emission components from the spectroscopic analysis of \citet{kowalski2013}.The transition time would correspond to the time when the two regions were equal in luminosity.

Likely neither of these simple parameterizations is entirely correct for describing the evolution of white light emission region(s) in a flare. Thus, we have provided both sets of equations for use in comparing to future flare atmosphere models. For the remainder of our paper we adopt the latter parameterization, given in Equation 4.

\section{Complex Flares}
In our final, by-eye validated sample of over 6000 flare events, more than 15\% were classified by users as ``complex''. Observationally this meant a significant secondary peak was present in the light curve before the flux returned to the quiescent level. While our automatic flare-finding algorithm attempted to quantitatively define which events were ``complex'', as described in \S2.1, the ultimate choice was determined by the average classification selected by the users. This subjective classification was biased towards recovering complex events with large secondary peaks that are well separated from the primary peak. Conversely, when the secondary structure has a small amplitude, or is close to the primary peak, human classification tended to be less accurate. In this section, we seek to produce a quantitative, objective classification for complex flare events using the flare template developed above.

\subsection{Fitting Complex Flares}
The analytic (classical) flare template we have developed can be used as a model or basis function to decompose complex events. In this ansatz, complex events are described as the superposition of several classical flares. Linearly adding a series of our models then reproduces the observed complex morphology. For a given complex event, the task is to determine both the specific properties of each constituent flare, namely ($t_{peak}$, amplitude, $t_{1/2}$), as well as the {\it number} of classical flares needed to describe the event.  

\begin{figure}[]
\centering
\includegraphics[width=3.5in]{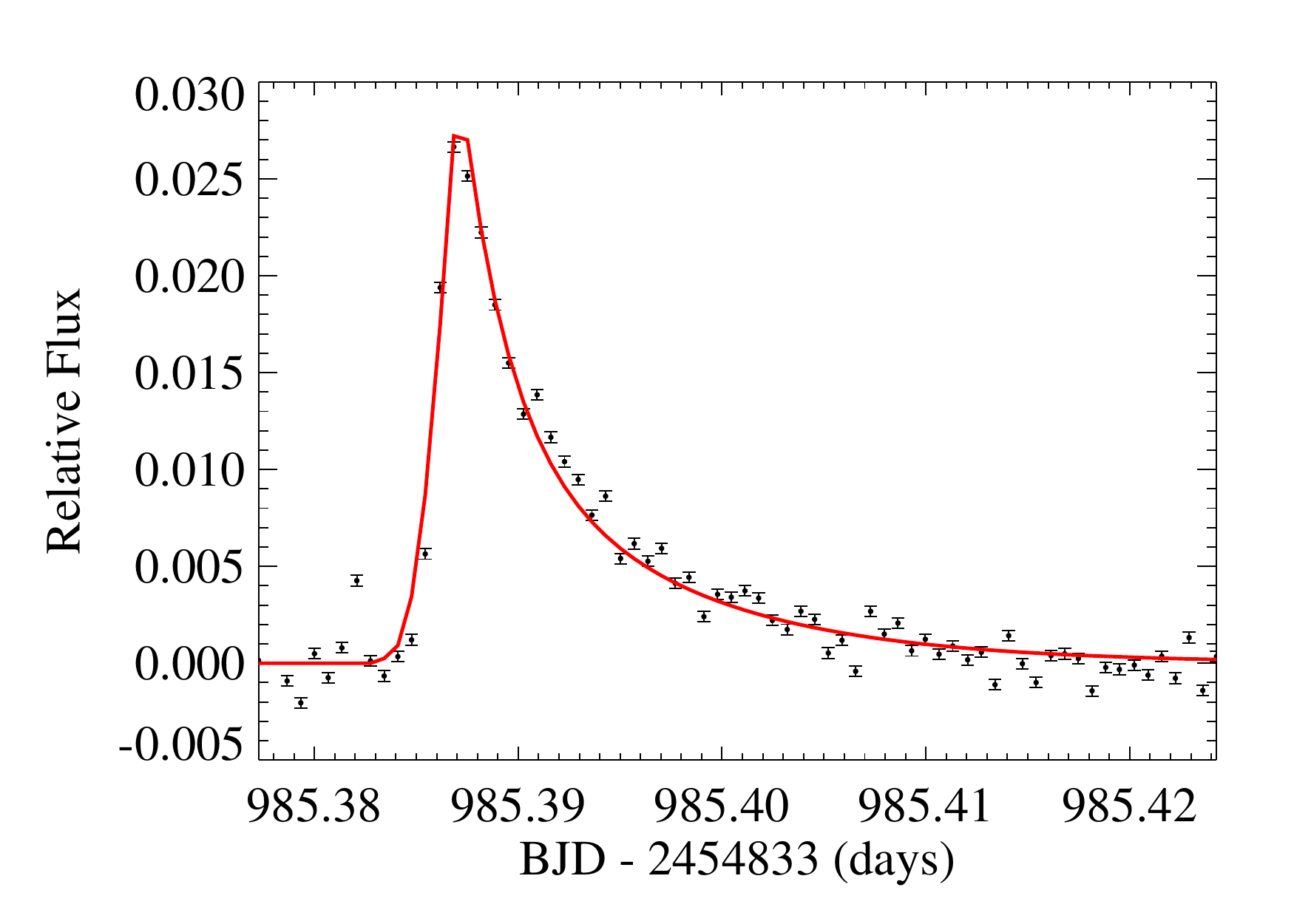}
\includegraphics[width=3.5in]{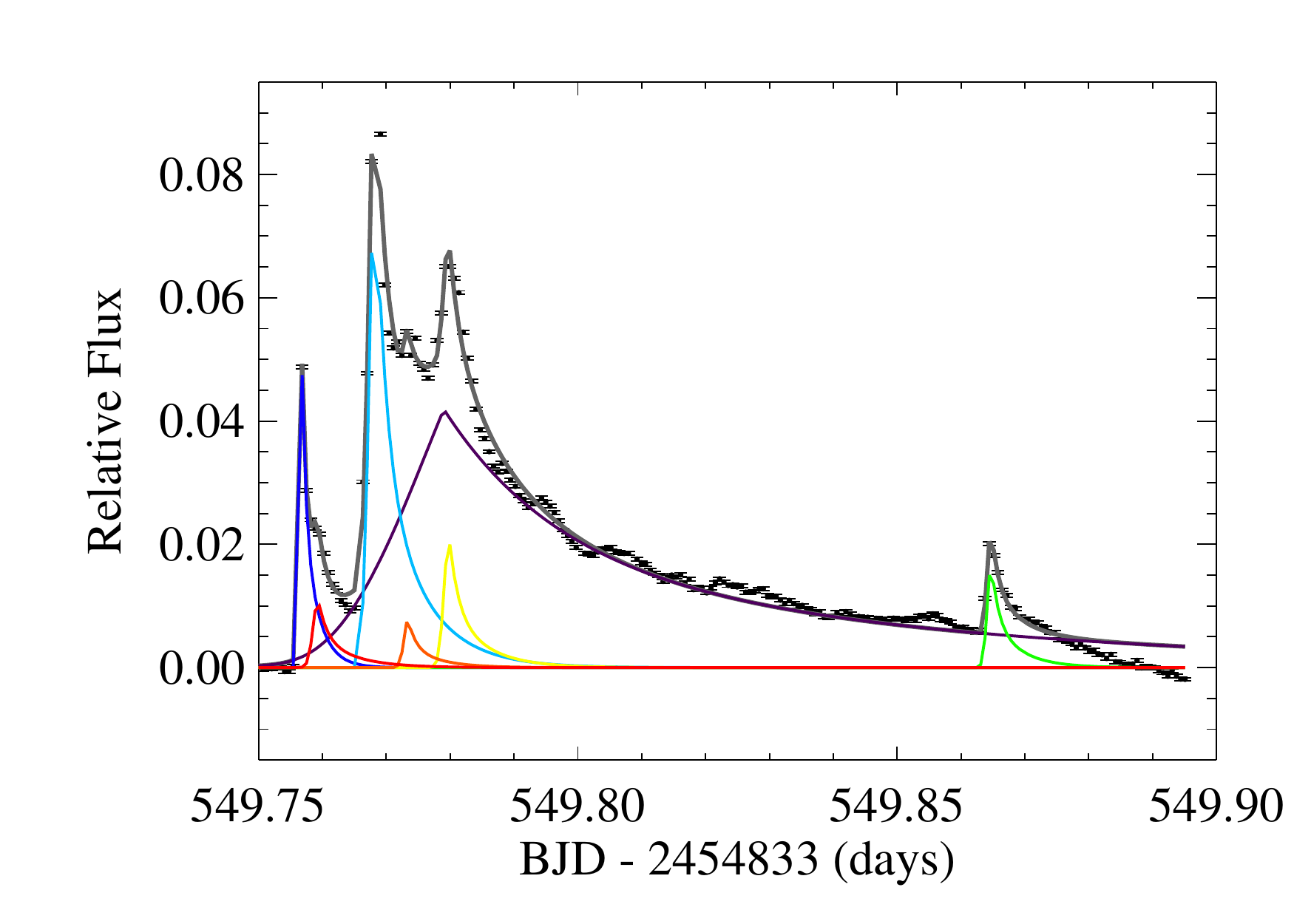}
\caption{Two examples of model fits to flare data. Top: a classical flare event that is well-fit by the template. Bottom: a complex flare event that required seven template flares to produce a good fit to the light curve.}
\label{cplxfit}
\end{figure}

We fit model flares to every event in the flare sample with a (by-eye determined) duration of 10 minutes or greater. This sub-sample included 3737 individual flare events. For each event, we fit 1 to 10 individual flare models. This was done iteratively, and the final ``best-fit'' number of flares was determined using the Bayesian Information Criterion. Two example flare events fit by this procedure, which we describe in detail below, are shown in Figure \ref{cplxfit}.

Every flare event was fit in fractional flux units relative to the global median flux (Figure \ref{fig:lc}), and using the de-trended starspot-removed light curve. The flare region to fit for each event included 3 minutes before the user-defined $t_{start}$ and 15 minutes after $t_{stop}$, to ensure that the flare fully returned to the quiescent flux level within the fitting window.  As was done in constructing the median flare model, we subtracted any residual starspot signal using a linear fit to small windows of time before and after each flare region.

For the first pass ($n=1$ model) we fit each event with a single flare template. We again utilized the IDL non-linear least squares fitting suite MPFIT \citep{mpfit} to minimize the empirical model to the flare event. We seeded this minimization using the peak flux amplitude and time, and 15\% of the full duration for $t_{1/2}$. The resulting $n=1$ best-fit was subtracted from the observed event, and the maximum peak in the residual flux was assumed to be an additional flare. To seed the $n=2$ model minimization, we used the best-fit $n=1$ model values, as well as the amplitude and time of the largest positive flux residual, and again 15\% of the full event duration for $t_{1/2}$. The $n=2$ model was then minimized, allowing both the primary and secondary flare to be fully solved to the observed data. We repeated the process of fitting and subtracting for $n=1$ through $n=10$ models on each flare event, solving for every flare component in each iteration.

Boundary conditions on the fit parameters were also imposed. Each component flare amplitude was required to be larger than twice the average photometric uncertainty within the flare event. We also required $t_{peak}$ for each component to occur within the boundaries of the flare time window, and $t_{1/2}$ for each component to be larger than 1 minute and smaller than 50\% of the observed flare duration. No priors on the relationships between flare amplitude, $t_{1/2}$, and $t_{peak}$ were included in our fitting.

We then choose the ``best'' model to represent each event using the Bayesian Information Criteria (BIC). This statistic attempts to determine the improvement of the fit (decreased $\chi^2$) while penalizing the increasing number of free parameters used in subsequent models. We computed the BIC for all $n=1$ through $n=10$ models, which took the form 
\begin{equation*}
{\rm BIC}_n = \chi^2 + k_n \ln( M)\,,
\end{equation*}
where $M$ was the number of observed data points that fell within the flare event window, and $k_n$ the number of degrees of freedom in the $n$'th model. The ``best fit'' solution was then selected as the $n$'th model with the smallest BIC parameter, where we additionally required the BIC to have decreased by at least 10\% from the previous ($n-1$) model. Finally, complex flares were defined to be any event best fit with a $n>1$ model. The choice of a 10\% BIC improvement threshold was determined by manual inspection of repeated fits to complex flares, and ensures we are not over-fitting these events.

In Figure \ref{cplxfit}, we show two examples of this fitting procedure. The top panel demonstrates a classical flare, very well fit by the $n=1$ solution shown in red. The bottom panel contains an example
of a very complex event, with four clearly separated peaks in the light curve. Our best model decomposed this event in to 7 distinct components, shown individually as colored lines. While the total morphology of this event is well fit by our procedure, a few discrepancies are apparent. For example, the broadest component flare (purple) does not account for the 4 or so low-amplitude secondary flares in the gradual decay phase. Additionally, the $t_{peak}$ for this component aligns with a much smaller amplitude component (yellow), which may not be physically realistic. This broad component flare over-estimates the flux near the end of the flare event, further indicating that this component should have a shorter timescale, and additional component flares should be fit in the decay phase. This example illustrates both the utility of our template in decomposing these events and classifying them as complex, as well as the intrinsic degeneracies in such a procedure.

\subsection{Rate of Complex Flares}

\begin{figure}[]
\centering
\includegraphics[width=3.5in]{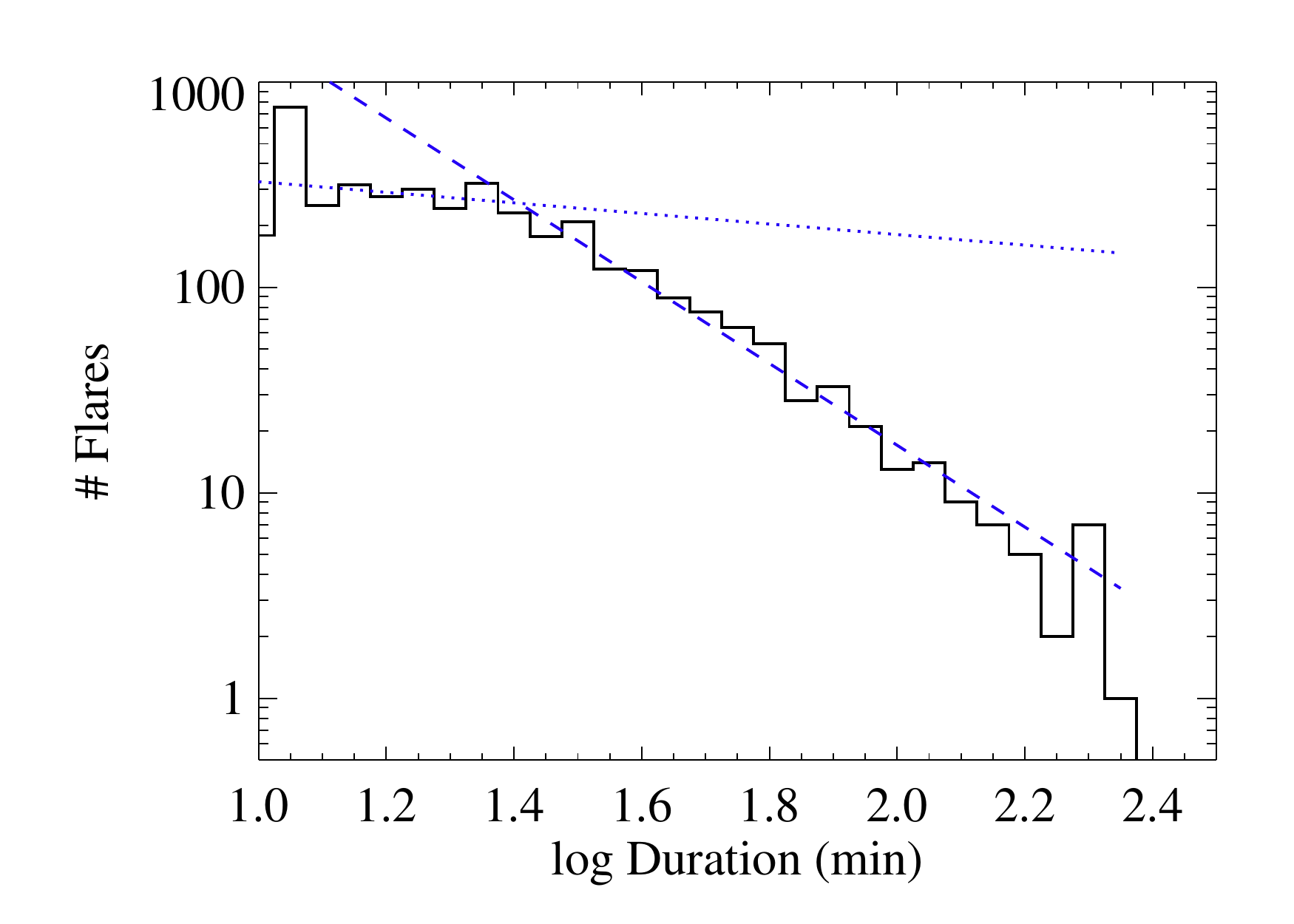}
\includegraphics[width=3.5in]{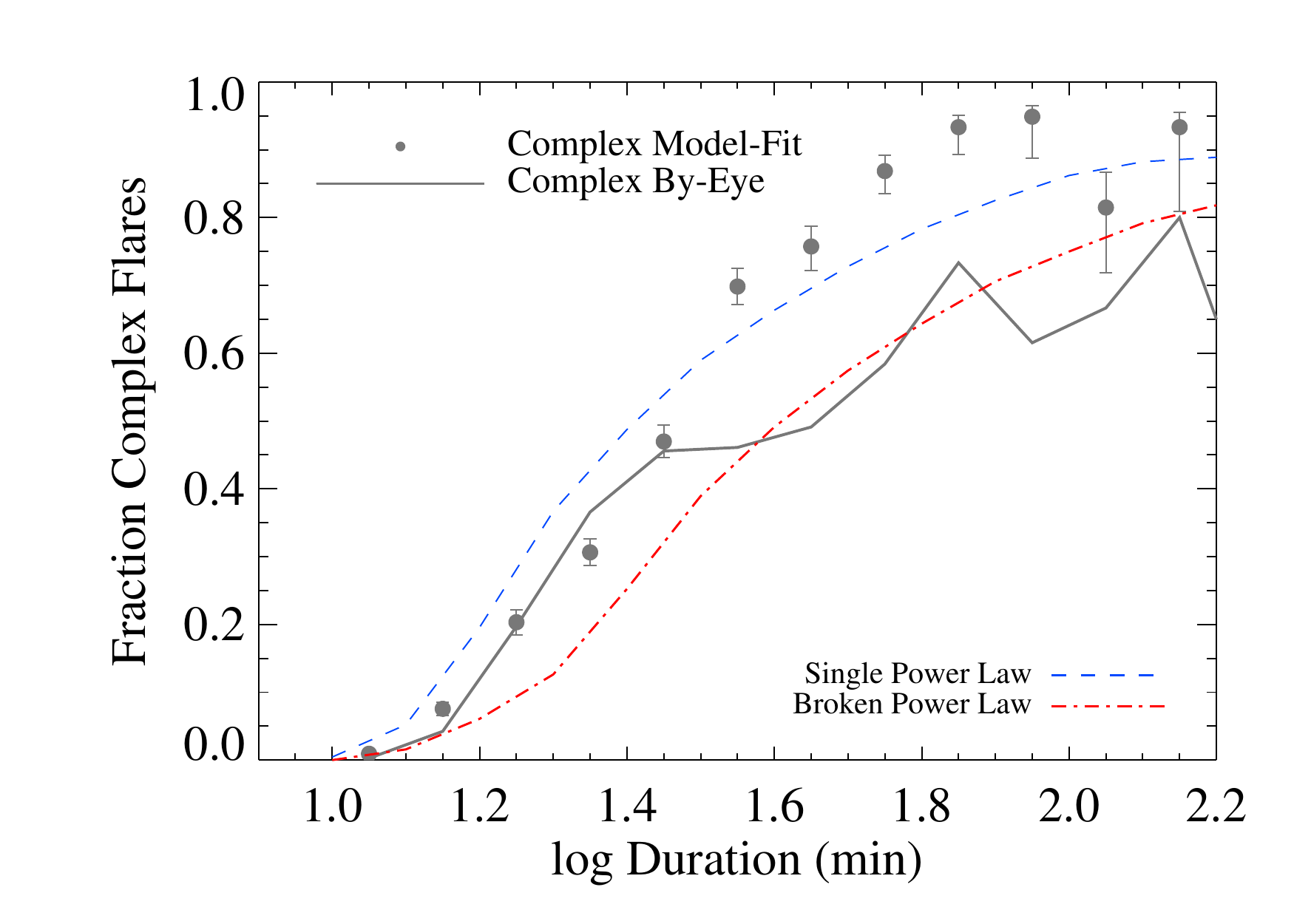}
\caption{Top: Distribution of flare durations for all events longer than 10 minutes in our sample. Single power law (blue dashed line) and broken power law (blue dashed and dotted lines) fits are shown. The total event duration as selected in our by-eye analysis in \S2 was used for each event. Bottom: Fraction of flares identified as complex using our iterative flare-fitting technique (black circles with error bars, see text for description), compared to those selected in our by-eye analysis (grey solid line). Two Monte Carlo models are overlaid (red and blue dashed lines), corresponding to the single and broken power-law durations distributions, see text.}
\label{cplx}
\end{figure}

We have thus far described the light curve structure of complex flares using superpositions of our empirical flare template. This methodology implies the varied substructure seen in complex events is the result of additional flares, which have the same morphological properties as classical events. From this approach two physical interpretations for complex flare events are possible: either 1) the multiple component-flares seen in complex events are physically associated within a single or nearby active regions on the star, or 2) complex events occur due to random superpositions of unassociated flares from separate active regions on the star. The former interpretation may include phenomena such as homologous flares, induced or sympathetic flares, and large tangled or multi-loop structures, all of which are seen on the Sun. Large complex flare events on M dwarfs, such as the one in \citet{kowalski2010}, have previously been interpreted in this manner \citep{anfinogentov2013}. Given the high level of magnetic activity and flaring on GJ 1243, the latter interpretation that some complex flares are chance superpositions likely accounts for at least {\it some} of the observed complex events. Larger energy, longer duration flares clearly have a higher likelihood of overlapping other non-associated flare events. 

Our user-validated flare sample included a flare classification of complex versus classical, as described in \S2.2. However, deciding if an event is complex requires that the two (or more) underlying flares be sufficiently separated in time, at least by a few minutes, such that the peaks are distinct. Additionally, the amplitudes of secondary flares within complex events must be high enough to clearly be distinguished above the morphology of the primary flare. These two considerations mean that the complex flare rate determined by eye is a lower limit. In Figure \ref{cplx} (top) we show the distribution of flare durations for all events in our by-eye validated sample. The fraction of events classified by eye as complex as a function of their duration is presented in Figure \ref{cplx} (bottom, solid grey line). As naively predicted above, the rate of complex flares does increases with the duration.

As described in \S4.1, we fit all 3737 flare events with durations of 10 minutes or longer with $n=1$ through $n=10$ template fits. This produced a sample of 1141 flares (30.5\%) that were best fit by an $n>1$ model, as in Figure \ref{cplxfit} (bottom), which we then classified as complex events. The fraction of model-fit complex events as a function of duration is also shown in Figure \ref{cplx} (bottom, black points). The uncertainties were computed from the binomial error within each duration bin. The rate of complex flares from the model was higher than in our by-eye sample, as expected.  Flares with durations greater than $\sim$50 minutes (log duration $\sim$1.7) show more than 80\% chance of being complex events from the model, but only 60\% from the users.

To compare with the observed fraction of complex flares as a function of duration, we generated two Monte Carlo models for creating complex flares in our data. In the first model, we generated random flares with durations 10 minutes or greater, drawn from a single power law with slope $-1.99\pm0.02$, which was determined by fitting the observed flare duration distribution for events longer than 20 minutes, shown in Figure \ref{cplx} (top, dashed blue line). A single power law fit for flare energy distributions is commonly used (see Paper 1). In the second model, we used a broken power law, with a slope of $-0.25\pm0.03$ for flares with duration between 10 and 20 minutes, and $-1.99\pm0.02$ for flares greater than 20 minutes to more closely fit the data, as shown in Figure \ref{cplx} (top, dashed and dotted blue lines). This two component, broken power law model is similar to the flare energy distribution model of \citet{kashyap2002}, though we emphasize we do not extend our simulation to unobservable ``microflares''.

The procedure for both the single and broken power law Monte Carlo models was the same. We began each model with a series of 50 trials, simulating a fixed number of flare events with durations ranging between 10 and 200 minutes. For these 50 trials, we adjusted the number of simulated flares between 3000 to 7500 events, with the first trial having 3000 flares and each subsequent trial increasing the number of flare events by 90. In each trial the specified number of flares was drawn from the respective duration distribution, and then placed at random start times throughout a blank light curve with the same time sampling as our dataset.  Any flare events that overlapped in time were combined, and the resulting event was classified as complex. We saved the resulting total number of events, both complex and classical, for each trial. 

We used these 50 trials to determine the number of simulated flares required to reproduce the observed total number of flare events. A second order polynomial was fit to the number of resulting versus number of simulated flares. For these models we required the resulting number of flares to match observations for the number of events (both classical and complex) with durations of 20 minutes or greater, which included 1750 flares from our by-eye sample. For the single power law model, generating this number of $\ge$20 minute events required an input of 6587 simulated flares, while the broken power law model required 4147 flares.

Using the respective number of input flares needed to generate the observed number of $\ge20$ minute events, we repeated this procedure 1000 times for both models. The fraction of complex events as a function of the resulting flare durations were recorded for each of the 1000 trials. The averaged complex flare fractions for both the single and broken power law models are shown in Figure \ref{cplx} (bottom, blue and red dashed lines, respectively).

Our data appear to rule out a single power law for the underlying duration distribution. While the single power law model more closely reproduced the template-fit complex flare fraction curve, and was tuned to match the number of flares with durations 20 minutes or greater, it required far more flares than were seen in our data. The total number of short duration events (less than 20 minutes) resulting from this model exceeded our observation by over 1000. According to Paper 1, flares with durations between 10 and 20 minutes would be expected to have energies of log E $\approx$ 30.5--31 erg, and are easily detectable in the Kepler data for GJ 1243.

Instead, a broken power law is favored by our observations. Using the slopes fit from the observed duration distribution in Figure \ref{cplx}, the broken power law model was able to reproduce the user-selected complex flare fractions. This model, however, somewhat under-produced the 10--20 minute duration flares as compared to our sample. However, we have made very simple assumptions about the duration distribution. The true, underlying power law slopes for the distribution of individual flare durations are almost certainly different than what we observe, as many of the flares overlap and thus skew the observed duration distribution from its true shape.

The preferred (broken power law) model did not fully reproduce the the model-fit complex fraction. We believe this indicates that some of the complex structure our model-fitting scheme recovered must come from actual sympathetic flaring, for example from the same or other nearby active regions. We note that for a given complex event, from the Kepler observations alone we cannot differentiate whether the complex structure was due to random superpositions or sympathetic flaring. The overall rate of sympathetic flaring is likely represented by the excess complex flare fraction above the broken power law model. Observations with higher time cadence, and/or bluer wavelength coverage, may be able to detect additional lower amplitude complex flare structure, particularly for shorter duration flares. Further study on the complex flare fraction for other active stars would also be useful to constrain these model results. Studies of less active stars, where random superpositions are less frequent, will also help constrain the rate of sympathetic flaring.

\section{Sample Completeness}
As the sample of flares defined in this work is the largest such dataset for a single active M dwarf, our flare catalog will be a benchmark for many future empirical and theoretical investigations. Our flare auto-finding methodology is also generally useful for single-event detection in light curves. As such, it is critical that we accurately characterize the completeness and limitations of our sample. 

We first quantified the completeness of the auto-finder algorithm by injecting artificial flares in to our light curves and testing for their recovery. A total of 500,000 artificial flares were tested in our 11-month dataset. 
Each artificial flare was generated using the flare template described by Equations 1 and 4. The $t_{1/2}$ and amplitude parameters for each artificial flare were drawn randomly from ranges of 1--60 minutes and 4\e{-5}-- 4\e{-2} relative flux, respectively.  The flare peak-time was then placed at a random time within the 11-month light curve.

\begin{figure}[t]
\centering
\includegraphics[width=3.5in]{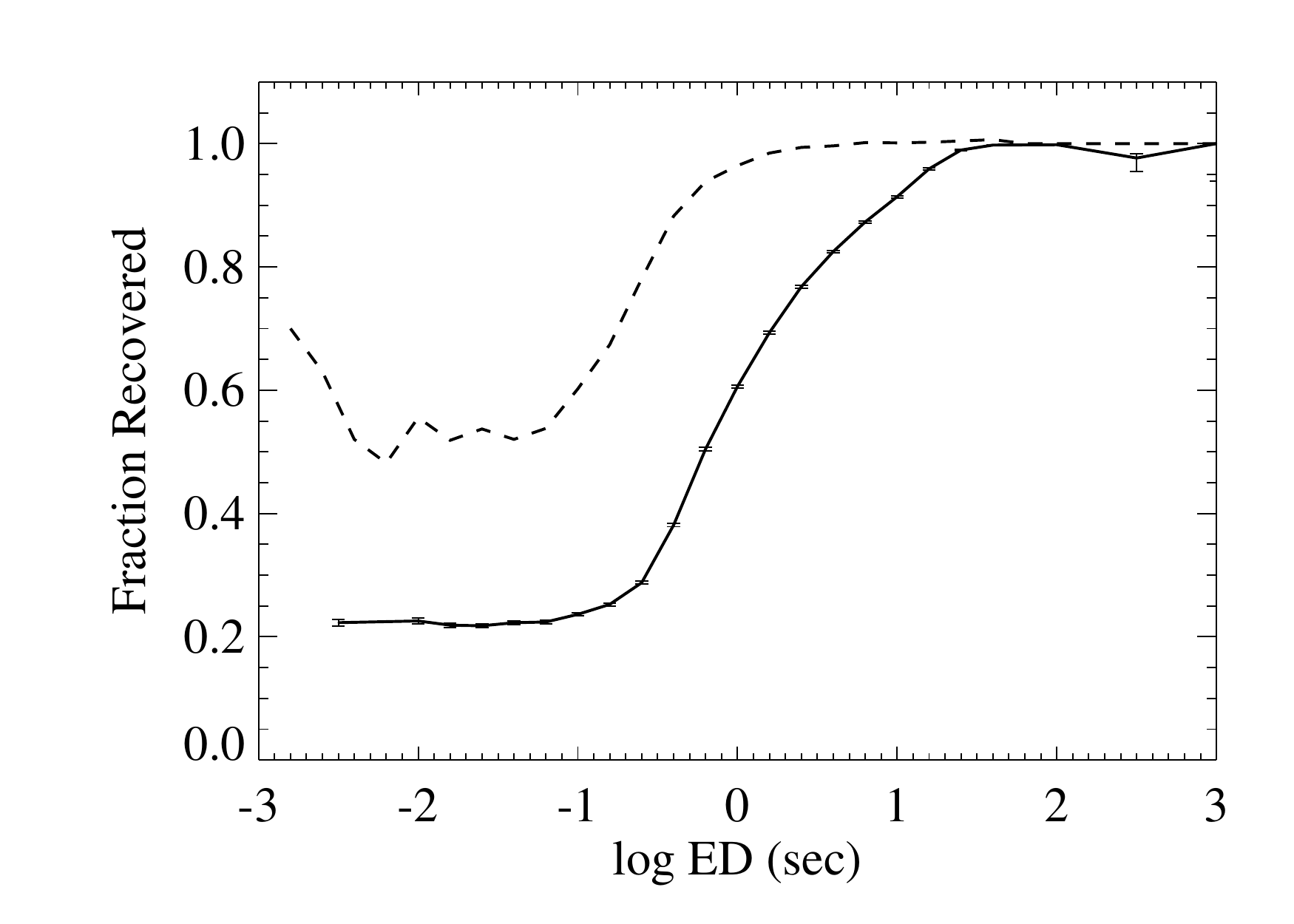}
\caption{Median fraction of recovered flares by the automatic flare-finding algorithm from tests using 500,000 artificial flares, calculated as a function of the event energies (solid black line). Error bars on each bin represent the binomial errors on flare recovery. Our method is 90\% complete for flares with ED greater than 10 seconds, and 70\% complete for flares with ED greater than 1 second (approximately 10 minute duration events). For comparison, the fraction of users who confirmed the flares in our final sample is shown for the same bins of event energy (dashed line).}
\label{fig:complete}
\end{figure}

Flares were considered recovered if the peak time of the input artificial flare was contained within any flare event found by the auto-finder. Care was taken to prevent the input artificial peak times from overlapping known flare events in the light curve. However, artificial flares that were placed near real large flares in the light curve could be combined into a complex event by the auto-finder. These were also considered recovered events. We did not test recovery of the input flare duration or amplitude. In Figure \ref{fig:complete} (solid line) we show the fraction of recovered flares as a function of their event energies (in units of equivalent duration). The median fraction of recovered flares was computed in bins of log ED = 0.2 sec. The auto-finder alone recovers events with log ED $>0$ sec, corresponding to flares with durations of 10 minutes (Paper 1), at 70\% completeness.

Our final flare sample was selected by a two-step procedure, first by the automated detection and then refined by human validation. As such, measuring completeness of the sample is a difficult task, since it would require measuring both the performance of the auto-finder, and studying the agreement between humans. This could theoretically be accomplished by injecting artificial flares into the light curves and tracing their recovery in both the auto-finder and the human validator steps, but would require a severe increase in human labor that was not practical. We therefore look at the user results separately.

In Figure \ref{fig:complete} (dashed line) we show the fraction of users who selected the flares in our final vetted sample as a function of their energies. Every flare in our sample was checked by at least five users. We used the peak time for each flare to compute the number of users who selected the event, and normalized by the number of users who validated the respective month of data. The median fraction of users who identified flares was again computed in bins of log ED = 0.2 sec. Users identified flares with 90\% agreement for events with log ED $> -0.5$ sec, corresponding to flares with durations of  $\sim$5 minutes (Paper 1). This illustrates that our final sample is 90\% complete, and that the auto-finding methods, while useful as a first pass, still need additional work to approach the confidence level of identifying by eye.

\section{Unusual Flares}
We have shown that our flare template is able to reproduce the morphology for many types of flare events, both classical and complex. However, a small fraction of flare events in our sample were not well fit by our template. To identify unusual flares in our sample, we selected events whose best-fit model had a  reduced $\chi^2$ of at least 15. This yielded 49 events (1.3\%) that were poorly fit by our iterative model approach. These were preferentially higher energy flares: median log E = 32.4 erg for unusual events compared to median Log E = 30.8 erg for the entire sample of flares of duration $\ge$10 minutes. We note that the rate of unusual flares is highly dependent on the details of our model fitting scheme, and changes to the BIC criteria or the iterative approach may result in fewer flares being classified as unusual.

Our model is able to fit events classified as ``impulsive'' and ``hybrids'' by \citet{kowalski2013}. Fitting their ``gradual'' flares (specifically GF1) would require either a large number of small component flares or a different template entirely. Without multi-wavelength or spectral data for such events, we are unable to address the underlying physical differences between these unusual or gradual flares, and the majority of the impulsive flares that are similar to our template.

\begin{figure}[t]
\centering
\includegraphics[width=3.5in]{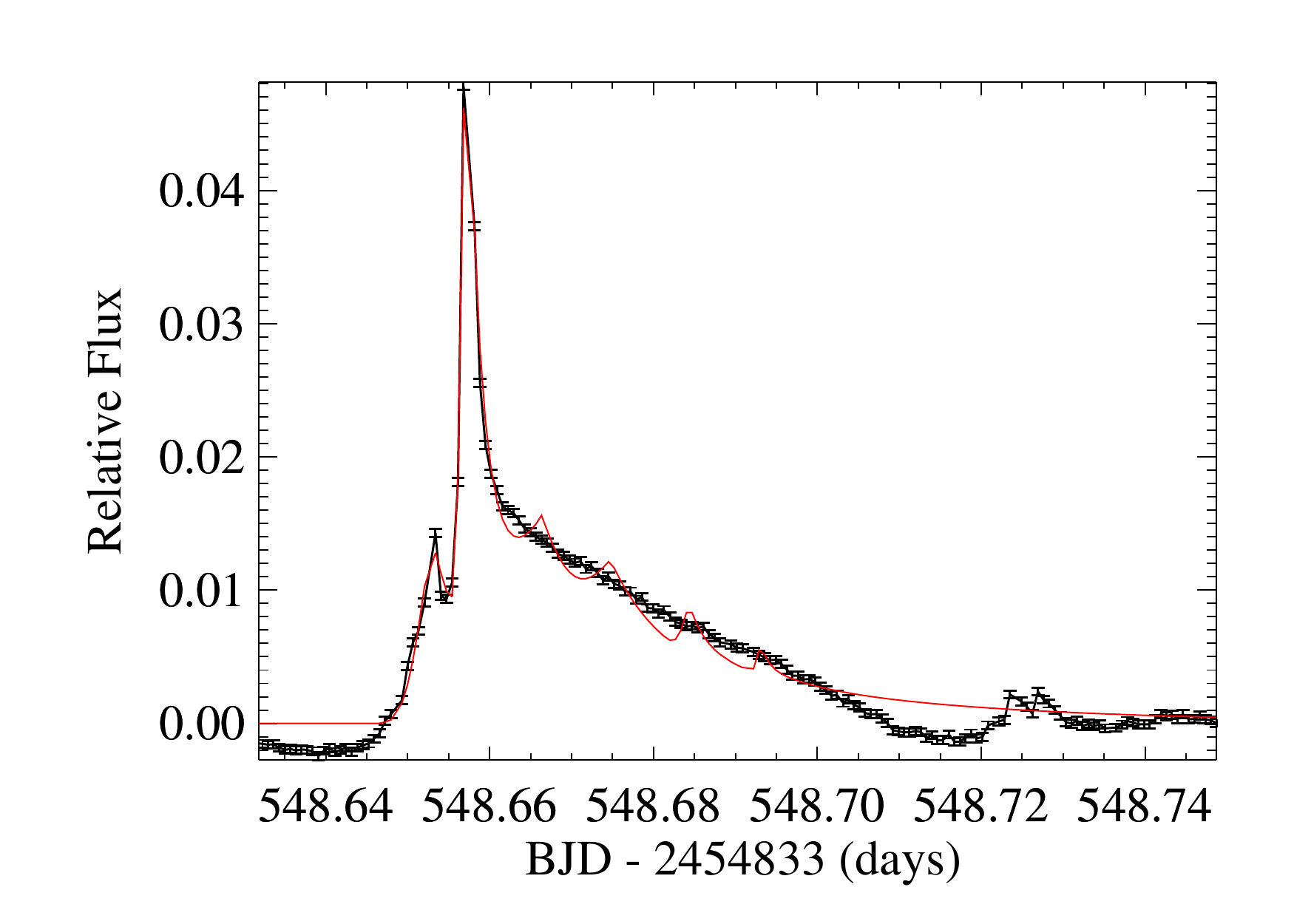}
\caption{An example of a flare with an unusual decay phase profile that is not well-fit by a combination of classical flare templates (black points). The best-fit model for this event (red line) included six component flares, but does not reproduce the structure observed in the decay phase. Negative flux before and after the event was an artifact of over-subtraction of the starspot signal, and did not affect our flare model fitting.}
\label{weird}
\end{figure}

In Figure \ref{weird}, we show an example of a high-energy complex flare that was poorly fit by our model. The best-fit complex flare model, which had six component flares and a reduced $\chi^2=15$, is shown. Negative relative flux near the beginning and end of the flare is an artifact from subtracting the local starspot modulation using a linear fit, and does not affect our analysis. The two main peaks in the light curve are well fit by two impulsive flare components. The slow decay phase, however, does not follow the model template (see Figure \ref{fig:medflare}). Instead, after the largest amplitude peak, this flare exhibits a very slow, almost linear decay in flux for over an hour. Our model attempted to reproduce this decay with a series of decreasing energy impulsive flare events, seen as the subsequent ripples in the model light curve.  While such ripples have been observed in the decay of complex events \citep[e.g.][]{kowalski2010}, they do not follow the data for this flare. 

Manually fitting the long decay using a single low-amplitude and large $t_{1/2}$ flare component also could not reproduce the linear flux decay observed.
The gradual decay profile of this flare could be reproduced using a much larger number of short timescale flares with decreasing energies. This is reminiscent of flare models used to reproduce the gradual soft X-ray decay in Solar flares \citep{warren2006}, or of large flares from reconnection along arcades of loop structures \citep{grigis2005}. Such an unusual decay profile might also be produced as a result of a flare occurring near, and possibly rotating over, the stellar limb \citep{tovmassian2003}. Given the rapid rotation rate for GJ 1243 this would appear a reasonable possibility for long duration flares, though it should not affect the majority of flares in our sample. We may speculatively ascribe the evolution of the flare in Figure \ref{weird} to an event initially occurring relatively close to the limb of the star, which then rotated partially or fully out of view.


\section{Discussion}

We have presented the largest sample of flares ever compiled for a single star besides our Sun, totaling 6107 unique flaring events from over 11 months of photometric monitoring. Of these, 15\% were classified as complex flares. In Paper 1 we described the general properties of flares on GJ 1243, including correlations between the decay time, duration, and flare energy for both complex and classical flares, and the waiting time distributions of the flares. Paper 1 found no correlation between the rotation phase of the star and the flare rate or average flare energy. We note here that the ratio of complex to classical flares also is not correlated with the stellar rotation phase.

Using a subset of 885 well-measured classical (single peak) events, we  generated an empirical flare template, which we fit with a fourth-order polynomial rise and double exponential decay. This template has only two morphological parameters: the amplitude and the characteristic timescale, $t_{1/2}$. We demonstrated the utility of this template in decomposing complex, multi-peaked flare events into their constituent classical flares. This empirical template will be a powerful tool for flare investigations on other stars. Future studies will show whether the morphological properties observed in GJ 1243 flares (e.g. the impulsive versus gradual decay profiles) as well as the choice of a single characteristic timescale are generic for all stellar flares.

We have restricted ourselves to studying the flare morphology for a single star in the Kepler dataset, GJ 1243. A detailed spectroscopic and photometric characterization of this fascinating star is also underway (Wisniewski et al. 2014, in prep.) Modeling the starspot modulations in the light curve, which we have thus far treated as a noise source, has also produced constraints on the differential rotation rate, spot geometry, and spot lifetimes (Davenport, Hebb, \& Hawley 2014, in prep.)

The degree of self-similarity between the classical flares in our template sample is remarkable. The ``impulsive phase'' (rise and decay) of the flare dominates the characteristic timescale ($t_{1/2}$) used in generating the template. The slow decay phase, which contains roughly half the flare energy, is also well traced. This indicates that our single timescale parameterization is appropriate for flare data at this cadence and in this wavelength regime, and that the two cooling phases do not appear to be independent, at least for flares on this star. Higher cadence investigations will be useful to uncover the universality of this template shape for the rise and both decay phases.

However, as \citet{andrews1965} first noted, the shape of the slow decay phase may depend on the energy or temperature of the flare event. \citet{gershberg1973} described star-independent properties for the slopes of the decay phases of flares, using power-law decay shapes rather than the exponential functions we found in our template. Star-independent flare morphology was also described by \cite{shakhovskaya1989}. These results have been disputed by \cite{kunkel1974}, however. Building on the work in Paper 1, and using the methods developed in this paper, we intend to look at flares on stars across all spectral types in the Kepler database. This will address the possible dependence of flare light curve morphology on the underlying star, as well as characterize the predicted yield and properties of flares in next generation time domain surveys.

\acknowledgments
We gratefully acknowledge support for this work from NASA Kepler Cycle 2 GO grant NNX11AB71G, NASA Kepler Cycle 3 GO grant NNX12AC79G. SLH, JRAD and LH acknowledge support from NSF grant  AST13-11678. EJH, AFK, and SLH acknowledge support from NSF grant AST08-07205. JRAD wishes to thank Andrew C. Becker for valuable insights in model fitting, and John J. Ruan for discussions of time series analysis.

This paper includes data collected by the Kepler mission. Funding for the Kepler mission is provided by the NASA Science Mission directorate. Some of the data presented in this paper were obtained from the Mikulski Archive for Space Telescopes (MAST). STScI is operated by the Association of Universities for Research in Astronomy, Inc., under NASA contract NAS5-26555. Support for MAST for non-HST data is provided by the NASA Office of Space Science via grant NNX13AC07G and by other grants and contracts.


\end{document}